\documentclass[pra,aps,twocolumn,superscriptaddress,showpacs]{revtex4}

\usepackage{amsmath,amsfonts,graphicx,epstopdf}
\usepackage{epsfig}
\usepackage{amssymb}
\usepackage{bm}
\usepackage{setspace}
\usepackage{graphics}
\usepackage{cancel}
\usepackage{psfrag}
\usepackage{titlesec}

\begin{document}

\author{R.~Fermani}
\email{Fermani@ntu.edu.sg}
\affiliation{Nanyang Technological University, Division of Physics and Applied Physics, 21 Nanyang Link, Singapore 637371}
\affiliation{Centre for Quantum Technologies, National University of Singapore, 3 Science Drive 2, Singapore 117543}
\author{T.~M{\"u}ller}
\affiliation{Nanyang Technological University, Division of Physics and Applied Physics, 21 Nanyang Link, Singapore 637371}
\affiliation{Centre for Quantum Technologies, National University of Singapore, 3 Science Drive 2, Singapore 117543}
\author{B.~Zhang}
\affiliation{Nanyang Technological University, Division of Physics and Applied Physics, 21 Nanyang Link, Singapore 637371}
\author{M.~J.~Lim}
\altaffiliation{Permanent address: Department of Physics and Astronomy, Rowan University, 201 Mullica Hill Road, Glassboro, New Jersey, USA}
\affiliation{Nanyang Technological University, Division of Physics and Applied Physics, 21 Nanyang Link, Singapore 637371}
\author{R.~Dumke}
\affiliation{Nanyang Technological University, Division of Physics and Applied Physics, 21 Nanyang Link, Singapore 637371}

\title{Heating rate and spin flip lifetime due to near field noise \\
in layered superconducting atom chips}
\begin{abstract}
We theoretically investigate the heating rate and spin flip lifetimes due to near field noise for atoms trapped close to layered superconducting structures. In particular, we compare the case of a gold layer deposited above a superconductor with the case of a bare superconductor. We study a niobium-based and a YBa$_2$Cu$_3$O$_{7-\mathrm{x}}$ (YBCO)-based chip. 
For both niobium and YBCO chips at a temperature of 4.2 K, 
we find that the deposition of the gold layer can have a significant impact on the heating rate and spin flip lifetime, as a result of the increase of the near field noise. At a chip temperature of 77 K, this effect is less pronounced for the YBCO chip. 
\end{abstract}
\date{\today}
\maketitle

\section{Introduction}
\vspace{-0.1cm}
In the area of magnetic trapping of ultracold atoms, considerable attention has been recently devoted  to the interaction of atomic clouds with the surfaces of both superconducting atom chips  \cite{Scheel05,Haroche06,Skagerstam07,Scheel07,Ulrich07,Shimizu07,mueller08,Fortagh09,mueller09,Emmert09,Haroche09,Shimizu09} 
and superconducting solid state devices \cite{Tian,Lukin,Petrosyan,Verdu}. 
Technological advances will allow a new generation of fundamental experiments and applications involving the control of the interface between atomic systems and quantum solid state devices. The implementation of such technologies depends on the ability to control and efficiently manipulate atoms
close to superconducting surfaces. However, below a certain separation the atom-surface coupling is strong enough that the trapping potential is modified by the interaction of the atomic magnetic moment with the near-field magnetic noise. This leads to atom heating and spin flip induced atomic loss, which is the basic limitation in metal-based atom chips \cite{EXPATOMCHIP,HENKEL/99,FOLMAN&Al,SPINFLIP,Harber03}. The origin of near field noise lies in the  thermally-induced fluctuating currents and in the resistivity of the chip materials in accordance with the fluctuation-dissipation theorem. The magnetic field noise is significantly smaller in the vicinity of a  superconductor \cite{Scheel05,Skagerstam07}, however for the small distances involved, heating and thermally-induced spin flip may become relevant. 

The reduction of the spin flip rate has been shown experimentally for different chip types: with and without a gold layer above the superconductor \cite{Haroche09,mueller09,Shimizu09}. The purpose of the top gold layer in atom chips is to use them in mirror--MOTs \cite{Reichel99}. This can be achieved by coating the chip surface with a reflective metal film. The metal film can also protect the superconductor when it is operated near the critical current density, to avoid possible quenching \cite{mueller08}. However, the presence of the metal film above the superconducting layer may increase the near field noise significantly. 
\begin{figure}[h]
\begin{center}
\includegraphics[width=6.5cm]{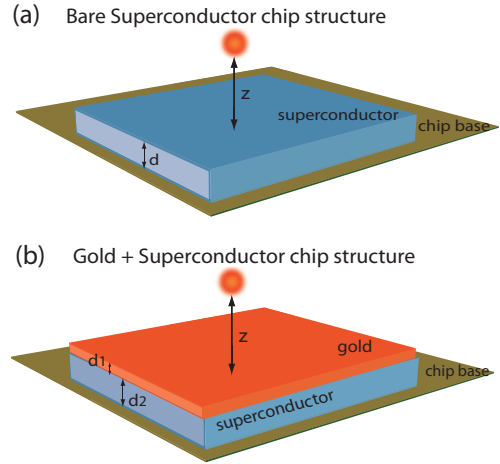}
\end{center}
\vspace{-0.6cm}
\caption{Schematic representation of the chip structure: (a) the bare superconductor chip structure consists of a layer of superconducting material of thickness $d$ deposited on a chip base (usually a crystal); (b) the superconductor+gold chip structure consists of a thin film of gold of thickness $d_1$ on top of the superconducting layer of thickness $d_2 < d_1$, which is deposited on the chip base. The superconducting layer is either niobium or YBCO.  }
\label{fig:st}
\end{figure}

We report our investigation of the heating rate and the thermally-induced spin flip rate of a neutral atom held close to two different superconducting structures: one with and one without a metal film above the superconducting layer. We consider the case of two different superconductors in the Meissner state: niobium (a conventional $s$-wave superconductor) and YBCO (a high temperature $d$--wave superconductor). We present a first-principles derivation of the heating rate, adopting a model for the heating mechanism similar to \cite{HENKEL/99,ghem}. For the derivation of the spin flip rate the reader may refer to earlier work \cite{SPINFLIP}. The evaluation of the rates is done within the framework of the electromagnetic field quantization in absorbing dielectric media \cite{VOGEL,PERINA}. Despite the quantum electrodynamics formalism being originally developed for dielectric and metallic media, it can be appropriately adopted to account for the electromagnetic field dissipation in superconducting materials in the Meissner state~\cite{Ulrich07}.

The paper is organized as follows. In Sec.~\ref{sec:model}, we describe the model we adopt for the heating rate and we give its general formulation. We present the quantum mechanical derivation of the heating rate in Appendix \ref{sec:qm}. In Sec.~\ref{sec:HR}, we report the formulas and simulations for the heating rate for two superconducting chip structures: with and without a metal layer. The comparison is done by considering niobium and YBCO superconductors. For completeness, we extend the comparison of the two materials by studying the spin flip lifetime in Sec.~\ref{sec:SF}. Conclusions are given in Sec.~\ref{sec:cl}.

\section{Model and Formulas}\label{sec:model}

\begin{figure}[ht]
\begin{center}
\includegraphics[width=5.8cm]{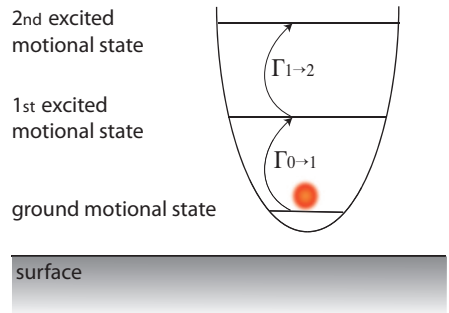}
\end{center}
\vspace{-0.6cm}
\caption{Schematic representation of the heating rate model. The heating rate is modelled as a transition from a given motional state of the trap to higher motional states. The trapping potential is considered  harmonic.}
\label{fig:sketch}
\end{figure}

The heating of an atom harmonically trapped can be understood as a transition from a motional state of the atomic trap to a higher motional state of the same trap. This is due to fluctuations of the trap center $\bm{x}$, as schematically represented in Fig.~\ref{fig:sketch}. 
We consider a three-dimensional harmonic oscillator with Hamiltonian 
$H= \bm{p}^2/2+\frac{1}{2}m \omega^2 \bm{x}^2 - \bm{x} \cdot \mathbf{F}$ where $\mathbf{F}$ is a force causing the trap centre to fluctuate. 
The force is proportional to the gradient of the magnetic potential experienced by the atom, which is a consequence of the Zeeman coupling of the atomic magnetic moment $\bm{\mu}$ to a spatially varying magnetic field $\mathbf{B}(\mathbf{r})$. 
Fluctuations of the magnetic field $\mathbf{B}(\mathbf{r}_A)$ at the atom's position $\mathbf{r}_A$ will induce fluctuations of the trap center and this process is described by the Hamiltonian 
\begin{equation}\label{eq:Hint}
\hat{H}_{int}= -\hat{\bm{x}} \cdot \nabla \left ( \bm{\mu} \cdot \hat{\mathbf{B}}(\mathbf{r_{A}})\right ),
\end{equation}
where $\hat{\bm{x}}=\sum_{j} x_j \hat{X_j}$, with $\hat{X}_j= (\hat{a}_j^{\dagger}+\hat{a}_j)/2$ being the dimensionless quadrature operator and $x_j$ the average position of the trap center along the $j$th direction. 
In the following, a concise notation will be adopted by dropping the sum over $j$ and assuming that $j=x,y,z$. 
For an atom initially in the state $|n\rangle$, the average trap center for a transition to a state $|m\neq n\rangle$ is given by $\bm{x} = \langle m| \hat{\bm{x}}|n\rangle$ which is nonzero only for $m=n \pm 1$, hence $x_{j,\pm}= \sqrt{2 \hbar/ M \omega_{v,j}} \sqrt{n+\frac{1}{2}\pm \frac{1}{2}}$, where $\omega_{v,j}$ is the trapping frequency along the $j$th direction and  $M$ is the atomic mass.

We obtain the rate 
$\gamma_{j,n\rightarrow n \pm 1}$ for the transition $n \rightarrow n \pm 1$ along the $j$th trapping direction as derived in Appendix~\ref{sec:qm}
\begin{equation}\label{eq:gamma}
\gamma_{j,n \rightarrow n \pm 1}=\frac{\pi x_{j,\pm}^2}{2\hbar^2} S_F ( \mathbf{r}_A, \omega_{v,j})\, , 
%=\frac{\pi}{M \omega_{v,i} \hbar} S_F (\mathbf{r}_A, \omega_{v,i})  \left (n+1/2 \pm 1/2 \right ).
\end{equation}
where $S_F ( \mathbf{r}_A, \omega_{v,j})$ is the spectrum of the fluctuating force causing the shift of the trap centre. 
In order to obtain the heating rate, we consider the average energy of the system $\langle E (t) \rangle =\sum_{n} P(n,t) \hbar \omega (n+\frac{1}{2}) $ where the sum is performed over the motional trap states $|n\rangle$, and $P(n,t)$ is the probability that at time $t$ the system is in the state $|n\rangle$. The average heating rate $\Gamma_{\epsilon,j} \equiv \langle \dot{E} (t) \rangle_j $ is given by \cite{ghem}
\begin{eqnarray}
\Gamma_{\epsilon,j} = \sum_n P(n,t) \hbar \omega_{v,j} \left ( \gamma_{j,n \rightarrow n+1}-\gamma_{j,n \rightarrow n-1}\right ),
\end{eqnarray}
and by substituting Eq.~(\ref{eq:gamma}) into the previous equation we obtain  
\begin{eqnarray}\label{eq:rateGen}
\Gamma_{\epsilon,j}=\frac{\pi }{M} S_{F}  (\mathbf{r}_A, \omega_{v,j}) \, , 
\end{eqnarray}
where $\sum_n P(n,t)=1$. 
The expression of  Eq.~(\ref{eq:rateGen}) is the most general formula we obtain for the heating rate and must be evaluated for each separate structure. In particular, the spectrum of the fluctuating force has been obtained in Appendix \ref{sec:qm} as 
\begin{eqnarray}
\lefteqn{
S_F (\mathbf{r}_A, \omega) = \frac{\hbar}{\pi \epsilon_0 c^2} (n_{\mathrm{th}}+1)}
\nonumber \\
&& \overrightarrow{\bm{\nabla}} \left ( \bm{\mu} \cdot 
\mathrm{Im} \left [\overrightarrow{\bm{\nabla}} \times  \bm{G}(\mathbf{r}_{A},\mathbf{r}_A,\omega) \times \overleftarrow{\bm{\nabla}} \right ] \cdot  \bm{\mu} \right ) \overleftarrow{\bm{\nabla}}\, , 
\end{eqnarray} 
where $n_{\mathrm{th}}= 1/(e^{\hbar \omega/k_B T}-1)$ is the mean thermal photon number and $\bm{G}(\mathbf{r}_{A},\mathbf{r}_A,\omega)$ is the Green function. The Green function accounts for the spectrum of the fluctuating magnetic field according to the following expression 
\begin{eqnarray}\label{eq:SB}
\lefteqn{S_B (\mathbf{r},\mathbf{r}', \omega) =\langle \hat{\mathbf{B}} (\mathbf{r},\omega) \hat{\mathbf{B}} (\mathbf{r}',\omega')\rangle}
\\
&=&  \frac{\hbar}{\pi \epsilon_0 c^2}   (n_{\mathrm{th}}+1)
\mathrm{Im} \left [\overrightarrow{\bm{\nabla}} \times  \bm{G(\mathbf{r},\mathbf{r}',\omega) \times \overleftarrow{\bm{\nabla}} \right ]} \delta (\omega-\omega').
\nonumber 
\end{eqnarray} 
All the information regarding the trapping parameters and the geometry of the system are contained in the Green function.

\section{Heating rate}\label{sec:HR}

We present our numerical evaluations for the heating rate of Eq.~(\ref{eq:rateGen}) for two different chip structures as shown in Fig.~\ref{fig:st}(a)-(b). Both structures contain a superconducting layer, either niobium or YBCO, in the Meissner state. We evaluate the heating rate for a $^{87}$Rb atom held at a distance $z$ above the chip. We consider the following trapping frequencies along the three directions in space: $\omega_{v,x}=0.1$ KHz and $\omega_{v,y}=\omega_{v,z}=1$ KHz.

The heating rate along the $j$th trapping direction for the niobium-based chip structure can be obtained from Eq.~(\ref{eq:rateGen}) by substituting the Green functions for isotropic media as reported in \cite{DUNG/98}, leading to  
\begin{eqnarray}\label{eq:rateS}
\lefteqn{
\Gamma_{\epsilon,j} = \alpha_j \frac{\mu_B^2 \hbar}{M \epsilon_0 c^2}   (n_{\mathrm{th}}+1)}
\\
&& \int \frac{ d\eta \, \eta^2}{4 \pi} 
\frac{e^{- 2\eta z} }{2}
  \mathrm{Im} \left [   R^{TM} \frac{\omega^2_{v,j}}{c^2}  +2 \eta^2 R^{TE}  \right ]\,  
  \nonumber 
 \end{eqnarray}
where $\alpha_x = \alpha_y=1$ and $\alpha_z=2$, and $\mu_B$ is the Bohr magneton. The Fresnel coefficients $R^{TE}$ and  $R^{TM}$, for $TE$ and $TM$ waves, are computed according to the nature and number of layers constituting the structure. In particular, the Fresnel coefficients for the $n$th layer follow the recursive formula
\begin{eqnarray}\label{eq:Fresnel1}
R_{n,m}^q = \frac{r_{n,m}^q +R^q_{m,m+1} e^{2 i \beta_{m} d_m}}{1+r_{n,m}^q R^q_{m,m+1} e^{2 i \beta_{m} d_m}}\, , 
\end{eqnarray}
where $m=n+1$ denotes the consecutive layer to $n$, with thickness $d_m$, $\beta_{m}= \sqrt{k_{m}^2-\eta^2}$, $k_{m}^2= \epsilon_{m} \omega^2/c^2$, and $\epsilon_m$ is the relative permittivity of the $m$th layer. The label $q$ denotes either the TE or the TM components of the electromagnetic field. The $r_{n,m}^q$ terms represent the Fresnel coefficients for the simplest geometry, two half spaces, and are given by 
\begin{eqnarray}\label{eq:Fresnel2}
r_{n,m}^{TE} = \frac{\beta_{n} - \beta_{m}}{\beta_{n}+ \beta_{m}}, \quad \quad 
r_{n,m}^{TM} = \frac{\epsilon_m \beta_{n} - \epsilon_n \beta_{m}}{\epsilon_m \beta_{n}+\epsilon_n \beta_{m}}.
\end{eqnarray}
\begin{center}
\begin{figure}[h]
\includegraphics[width=8.3cm]{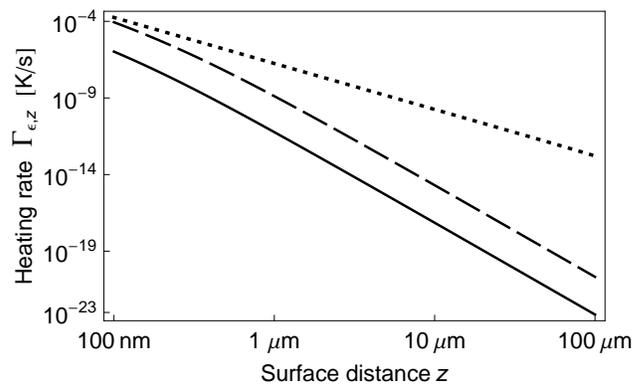}
\caption{Heating rate for a Nb-superconducting chip at 4.2 K, the thickness of the Nb layer is 1 $\mu$m. Each of the three curves represents a different structure: the bare niobium chip (solid line), the niobium+gold chip with gold thickness of 50 nm (dashed line) and a simple gold surface (dotted line). }\label{fig:heatingNB}
\end{figure}
\end{center}

In Figure \ref{fig:heatingNB} we plot the heating rate $\Gamma_{\epsilon ,z}$ along the $z$ direction versus the surface distance for three different chip structures at liquid helium temperature $T=4.2$ K. We scale the heating rate by the Boltzmann constant in order to facilitate comparison with relevant energy scales, however this should not be confused with the thermodynamic temperature.  

The heating rate is highest above the gold substrate, while the bare niobium surface yields the lowest rate. 
At distances of $1 \, \mu$m, we estimate the heating rate for the gold substrate to be of the order of $\Gamma_{\epsilon,\mathrm{Au}} \simeq 10^{-7}$ K/s which is a measurable effect. At the same distance, the heating for the niobium-based chip is reduced but in principle still detectable with a top gold layer of 50 nm as $\Gamma_{\epsilon,\mathrm{Nb}+\mathrm{Au}} \simeq 10^{-9}$. However, the rate for the bare niobium chip is too small to be of experimental relevance because $\Gamma_{\epsilon,\mathrm{Nb}} \simeq 10^{-12}$ K/s. 
Trapping distances below $1\, \mu$m are challenging because of the van der Waals attractive potential. However, for $z \sim100$ nm, the heating rate of the niobium+gold surface approaches the heating rate of the gold substrate. This indicates that at very small distances only the effect of the metal is relevant. 

For the YBCO-based chip structure, the Green function for anisotropic media as reported in \cite{Green} is adopted  and the heating rate along the $z$-direction (perpendicular to the chip surface) is given by 
\begin{eqnarray}\label{eq:Yz}
\lefteqn{\Gamma_{\epsilon, z}=\frac{\hbar \mu_B^2 }{M \epsilon_0 c^2} (n_{\mathrm{th}}+1)}
\\
&& \int\limits_{0}^{\infty} \frac{d \eta \, \eta^2}{4 \pi}  e^{-2 \eta  z}
\mathrm{Im} \left [ B_{N 2}^{11}  k_z^2+2 \eta^2 B_{M 1}^{11}
 \right ], \nonumber
\end{eqnarray}
with $k_z^2=\omega_z^2/c^2$, while the heating rate $\Gamma_{\epsilon, \|}$ for the plane parallel to the chip surface is given by the sum of the following expressions 
\begin{eqnarray}\label{eq:Yr}
\lefteqn{\Gamma_{\epsilon, r} =\frac{\hbar \mu_B^2 }{M \epsilon_0 c^2} (n_{\mathrm{th}}+1)}
\\
&&  \int\limits_{0}^{\infty} \frac{d \eta \, \eta^2}{4 \pi} 
 \frac{e^{-2 \eta z}}{4}   
\mathrm{Im} \left [  B_{N 2}^{11}  k_{r}^2+ 3 \eta^2  B_{M 1}^{11} \right ], \nonumber
\\
\label{eq:Yfi} \lefteqn{ \Gamma_{\epsilon , \phi}  =\frac{\hbar \mu_B^2 }{M \epsilon_0 c^2} (n_{\mathrm{th}}+1)}
\\
&& \int\limits_{0}^{\infty} \frac{d \eta}{4 \pi}  e^{-2 \eta z}
\mathrm{Im} \left [   B_{N 2}^{11}  k_{\phi}^2 +\eta^2 B_{M 1}^{11} \right ],  \nonumber
 \end{eqnarray}
 \vspace{-0.1cm}
 with  $k_{r,\phi}^2=\omega_{r,\phi}^2/c^2$.
The scattering coefficients $B_{M 1}^{11}$ and $ B_{N 2}^{11}$ play the same role of the Fresnel coefficients of Eq.~(\ref{eq:Fresnel1}) but account for the anisotropic properties of the YBCO layer. These coefficients can be obtained by~following~\cite{fermani09,Green}.
\begin{center}
\begin{figure}[ht!]
\vspace{-0.2cm}
\includegraphics[width=8.3cm]{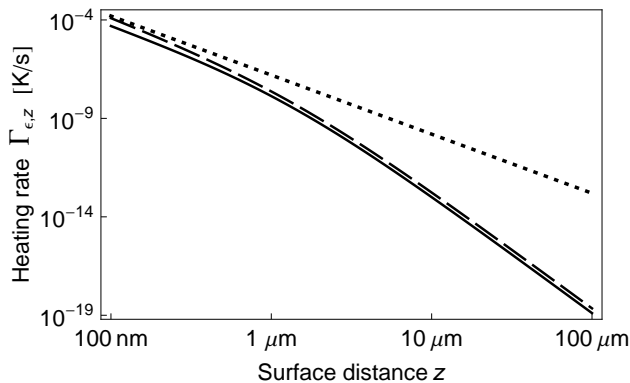}
\caption{
Heating rate for a YBCO-based chip at 77 K, the thickness of the YBCO layer is 1 $\mu$m. Each of the three curves represents a different structure: the bare YBCO chip (solid line), the YBCO+gold chip with gold thickness of 50 nm (dashed line) and a simple gold surface (dotted line).}\label{fig:heatingYB}
\vspace{-0.3cm}
\end{figure}
\end{center}

Our simulations for the heating rate of the YBCO chip are presented in Fig.~\ref{fig:heatingYB}. We plot the heating rate along only the $z$ trapping direction because this term dominates. We observe that the bare YBCO surface yields the lowest heating rate, compared to the other two structures with the gold layer. However, the difference between the bare YBCO surface and the YBCO+gold chip structure is dramatically reduced at liquid nitrogen temperature $T=77$ K, and is less than one order of magnitude for a wide range of surface distances. 
The heating rate is detectable for distances around $1\,\mu$m where $\Gamma_{\epsilon,\mathrm{YBCO}} \simeq \Gamma_{\epsilon,\mathrm{YBCO}+\mathrm{Au}} \simeq 10^{-8}$ K/s. At smaller temperatures, the rate corresponding to the two structures differs by a few orders of magnitude, i.~e.~$ \Gamma_{\epsilon,\mathrm{YBCO}+\mathrm{Au}} / \Gamma_{\epsilon,\mathrm{YBCO}} \simeq 10^{-3} $ at $T=4.2$ K given by the fact that the YBCO penetration depth and the skin depth of gold scale with temperature. This results in the near field spectrum being dominated by the gold. As a consequence, at very low temperatures the presence of the gold leads to a marked increase of near field noise. This increase is less pronounced at temperatures closer to the YBCO transition temperature. 
We conclude that the heating rate close to a bare YBCO surface and to a YBCO+gold chip for typical trapping distances of a few $\mu$m does not result in a measurable difference at $77$~K. Moreover, the typical values of magnetic bias fields and temperatures used for superconducting chip experiments suggest that a high-temperature superconductor is likely to be operated in the Shubnikov phase. This involves the partial penetration of magnetic field in the form of vortices but this is not considered in the present work.

\section{Spin flip lifetime}\label{sec:SF}
In the following section the comparison between the bare superconducting chip and the superconductor+gold structure is extended by considering the spin flip lifetime $\tau$ of a neutral atom.  The transition rate for thermally induced spin flips $\Gamma_{SF}=1/\tau$ between an initial hyperfine state $|i\rangle$ and a final hyperfine state $|f\rangle$ is given in its most general formulation as \cite{SPINFLIP} 
\begin{eqnarray}\label{eq:spin}
\Gamma_{SF} =  \frac{2 c \mu_B^2 g_S^2}{\hbar^2 } \langle f|\hat{S}_q |i \rangle \langle i |\hat{S}_p |f \rangle  S_B (\mathbf{r}_A, \omega_{SF}) \, , 
\end{eqnarray}
where $\omega_{SF}$ is the spin transition frequency and $\langle i |\hat{S}_p |f \rangle$ the spin matrix element for $|i\rangle \rightarrow |f\rangle$. 
We restrict the calculation to a two-level system evaluating the spin flip lifetime for the transition $|F=1,m_F=-1 \rangle \rightarrow |F=1,m_F=0\rangle$. We choose a typical experimental  transition frequency $\omega_{SF}= 2 \pi \, 560 $ KHz \cite{EXPATOMCHIP},  and we note that the near field noise spectrum does not change significantly in the \textit{rf} frequency range. We evaluate the spin matrix elements via the Clebsch-Gordan coefficients and obtain for the non-zero matrix elements $|\langle i | \hat{S}_{x}| f \rangle|=|\langle i | \hat{S}_{z}| f \rangle|=1/8$ (assuming that $y$ is the spin quantization direction).

The spin flip lifetime close to the niobium-based chip is obtained as the inverse of the following spin flip rate~\cite{fermani09}
\begin{equation}\label{eq:spinrateNb}
\Gamma_{SF,\mathrm{Nb}} =3 \pi  \frac{(\mu_B g_S)^2}{32 c^2 \epsilon_0 \hbar } \int \frac{d \eta \, \eta^2}{(2 \pi)^2} 
\frac{ e^{- 2\eta z} }{2 }  \mathrm{Im} \left [  R^{TE} \right ] (n_{\mathrm{th}}+1)\, , 
\end{equation}
where $R^{TE}$ is the Fresnel coefficient as in Eq.~(\ref{eq:Fresnel1}). 
The spin flip lifetime as a function of the surface distance is plotted in Fig.~\ref{fig:spinrateNb}  for three different structures. The shortest spin flip lifetime is obtained for the gold surface. The longest lifetime is found close to the bare niobium surface and is two orders of magnitude longer than the lifetime near the niobium+gold chip. 
\begin{center}
\begin{figure}[h!]
\includegraphics[width=8.3cm]{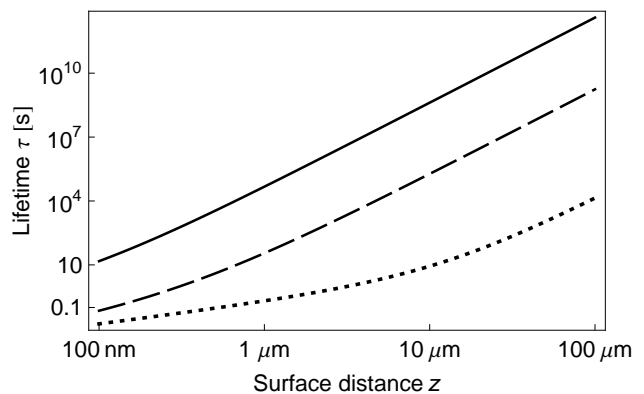}
\caption{Spin flip lifetime for a niobium-based chip at 4.2 K, the thickness of the Nb layer is 1 $\mu$m. Each of the three curves represents the lifetime corresponding to a different structure: the bare niobium chip (solid line), the niobium+gold chip with gold thickness of 50 nm (dashed line) and a simple gold surface (dotted line). } \label{fig:spinrateNb}
\end{figure}
\vspace{-0.6cm}
\end{center}

\begin{center}
\begin{figure}[ht!]
\includegraphics[width=8.3cm]{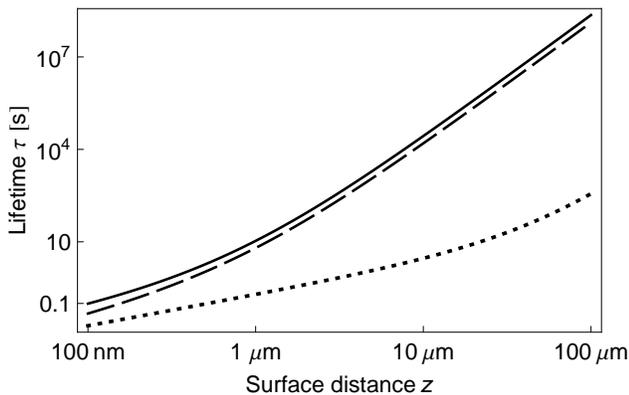}
\caption{Spin flip lifetime for a YBCO-based chip at 77 K, the thickness of the YBCO layer is 1 $\mu$m. The two upper lines represent the case with a bare YBCO wire (solid line) and with a gold layer of thickness 50 nm on top of YBCO (dashed line). The lifetime for a simple gold surface is reported for comparison (dotted line). } \label{fig:spinYB}
\end{figure}
\end{center}

The spin flip rate above the YBCO-based chip can be written as \cite{fermani09}
\begin{eqnarray}\label{eq:spinRateD}
\lefteqn{\Gamma_{SF,\mathrm{YBCO}}=\frac{ (\mu_{B} g_{S})^2}{32  \hbar \epsilon_0 c^2 }  (n_{\mathrm{th}}+1)}
\\
&&
\int\limits_{0}^{\infty} d \eta \, \frac{ e^{-2\eta z}}{8 \pi}
\mathrm{Im} \left [ 3 \eta^2 B_{M 1}^{11}+ 
B_{N 2}^{11} \frac{\omega_{SF}^2}{c^2}
\right ] , \nonumber
\end{eqnarray}
where $B_{M 1}^{11}$ and $B_{N 2}^{11}$ are the scattering
coefficients of Eqs.~(\ref{eq:Yz}-\ref{eq:Yfi}). The spin flip lifetime $\tau= 1/\Gamma_{SF,\mathrm{YBCO}}$ is plotted in Fig.~\ref{fig:spinYB} for the three different structures: gold surface, YBCO surface, YBCO+gold structure. 
Similarly to the niobium-based chip structures, the spin flip lifetime close to the
 YBCO+gold structure  is shorter than the one obtained near the bare superconductor, however the difference is less than an order of magnitude. 
A complete accounting of the total lifetime must include other experimental restrictions, such as collisions with background gas \cite{Fortagh09}. 
%%%%%%%%%%
%%%% 

\begin{center}
\begin{figure}[h!]
\hspace{-0.2cm}\includegraphics[width=8.1cm]{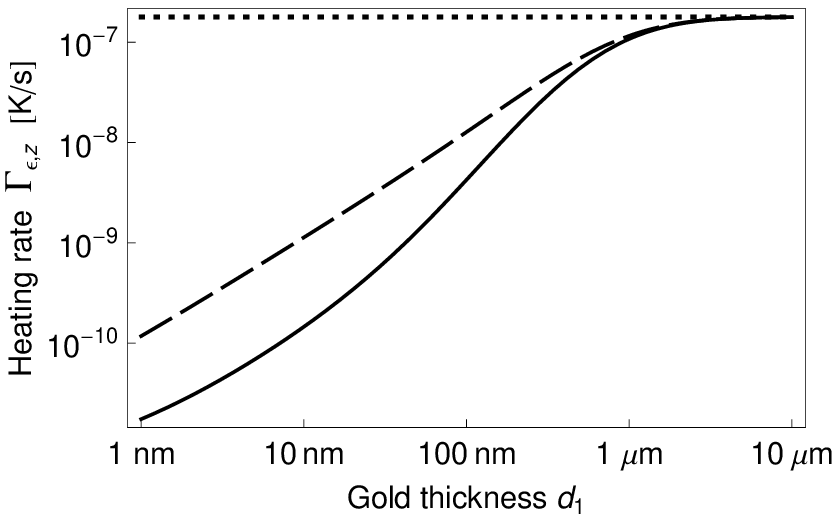}\\
\vspace{0.3cm}
\includegraphics[width=8cm]{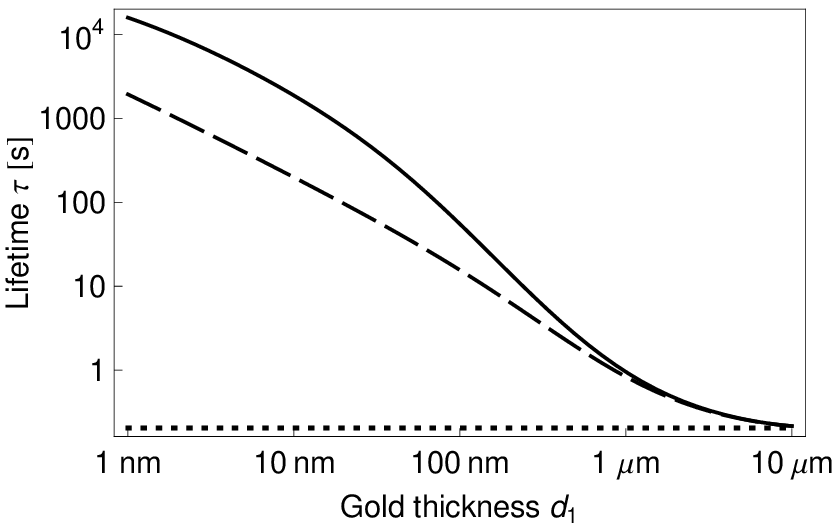}
\caption{
Heating rate and spin flip lifetime of an atom 1 $\mu$m above a superconducting chip with a top gold layer. Both quantities are plotted as a function of the gold thickness $d_1$ (as in Fig.~\ref{fig:st} (b)), for a niobium+gold chip (solid curve),  YBCO+gold chip (dashed curve) and for a gold substrate (dotted line). Both chips have a temperature of 4.2 K and the thickness of the superconducting layer is 1 $\mu$m.}\label{fig:tg}
\end{figure}
\end{center}

We conclude this section by plotting in Fig.~\ref{fig:tg} the heating rate and spin flip lifetime of the niobium+gold and YBCO+gold chip as a function of the gold thickness. 
The niobium+gold chip has the lowest heating rate. The rate increases for both structures with increasing the thickness of the gold until the layer thickness reaches $1\, \mu$m.  The presence of the superconductor is then irrelevant and the heating rate is the same as what one would obtain close to a gold substrate. Typical values for the gold thickness are around $50-100$ nm. For such values the heating rate for the YBCO+gold chip is one order of magnitude larger than the niobium+gold chip at 4.2 K. Similarly, the niobium+gold chip exhibits the longest lifetime. For both structures the lifetime decreases with increasing gold thickness until it approaches the lifetime obtained close to the single gold layer. This happens for a layer thickness of the order of $10 \, \mu$m.

\section{Conclusions}\label{sec:cl}
We have investigated the heating rate and spin flip lifetime for a superconducting atom chip considering the cases with and without a gold layer deposited above the superconductor. All the results presented are valid for a superconductor in the Meissner state. 
The heating rate for trapping distances around $1\, \mu$m is measurable and may become significant for a sufficiently cold atom cloud. However, by adding a top gold layer, this rate increases such that for layers thicker than $1\, \mu$m, the presence of the superconductor becomes irrelevant. Even for a sub-micron gold layer, an atom held at a close distance of $100$ nm is influenced only by the metal, and both the heating rate and spin flip lifetime approach the values obtained in the proximity of a infinitely thick gold substrate.

In particular, we have considered a niobium-based and a YBCO-based chip. At a chip temperature of 4.2 K, the difference between the case with the gold layer and without is marked both for heating rate and spin flip lifetime. Such difference is less pronounced when the YBCO chip is operated at 77~K. In summary, our study shows that the deposition of a metal layer above a superconducting layer can diminish the advantages of using a superconductor to overcome the limitations given by a metal. 
The presented results are of interest for future development of technologies involving superconducting atom chips. Further investigations will include the treatment of type II superconductors in the Shubnikov phase.

\acknowledgments
We acknowledge financial support from Nanyang Technological University (grant no. WBS M58110036), A-Star (grant no. SERC 072 101 0035 and WBS R-144-000-189-305) and the Centre for Quantum Technologies, Singapore. MJL acknowledges travel support from NSF PHY-0613659 and the Rowan University NSFG program.

%%%%%%%%%%%%%%%%%%%%%%%%%%%%%%%%%%%%%%%%%%%%%%%%%%%%%%%%%%%%%%%%%%%%%%
\appendix

\section{Quantum mechanical derivation of the heating rate}\label{sec:qm}

We outline the quantum mechanical derivation of the heating rate based on the model introduced in Sec.~\ref{sec:model}.

The total Hamiltonian of a harmonically trapped atom held close to a substrate (the conducting properties of the substrate are not yet specified at this stage), is given by the sum of three different Hamiltonians as
$ \hat{H}= \hat{H}_{f} +\hat{H}_{a}+\hat{H}_{int}$. 
The Hamiltonian for the atom in the unperturbed harmonic trap reads $\hat{H}_a=  \hbar \omega_{v,j} \left ( \hat{a}^{\dagger}_{j} \hat{a}_j + \frac{1}{2}\right )$. 
The Hamiltonian of the electromagnetic field arising from an absorbing and dispersing medium is given within the formalism of quantum electrodynamics for dielectric media \cite{VOGEL,PERINA} as 
$\hat{H}_{f}= \int d^{3}\mathbf{r} d \omega\, \hbar \omega\, \hat{\mathbf{f}}^{\dagger} (\mathbf{r},\omega)
\hat{\mathbf{f}} (\mathbf{r},\omega)$,
with $\hat{\mathbf{f}} (\mathbf{r},\omega)$ and $\hat{\mathbf{f}}^{\dagger} (\mathbf{r},\omega)$ being the bosonic operators accounting for the collective excitation of the medium and the electromagnetic field. They satisfy the usual equal-time commutation relations
$\left[\hat{\mathbf{f}}(\mathbf{r},\omega),\hat{\mathbf{f}}^{\dagger}(\mathbf{r}',\omega')\right]$
$\!=$ $\!\delta(\mathbf{r}-\mathbf{r}')\delta(\omega-\omega')$ and their correlation function at temperature $T$ reads 
$\langle \hat{\mathbf{f}}(\mathbf{r},\omega)
\hat{\mathbf{f}}^\dagger(\mathbf{r}',\omega') \rangle =
(n_{\mathrm{th}}+1) \delta(\mathbf{r}-\mathbf{r}')
\delta(\omega-\omega')$,
where $n_{\mathrm{th}}$ is the mean thermal photon number.%

The interaction Hamiltonian of Eq.~(\ref{eq:Hint}) can be written in the rotating wave approximation as 
\begin{equation}
\hat{H}_{int} = - \frac{x_j}{2}  \left (  \hat{a}_{j}^{\dagger} \partial_j  \mu_k \hat{B}^{+}_{k}(\mathbf{r_{A}}) +\mathrm{H.c.} \right ) \, , 
\end{equation}
where $x_{j}$ denotes either $x_{j,+}$ for $n \rightarrow n+1 $ or $x_{j,-}$ for $n \rightarrow n-1$. The magnetic field $B_{k}^{+} (\mathbf{r_{A}})$ can be obtained via
$\hat{ \mathbf{B}} (\mathbf{r} ) = \hat{\mathbf{B}}^{+} (\mathbf{r}) +\hat{\mathbf{B}}^{-} (\mathbf{r})$ where
$\hat{ \mathbf{B}}^{+ (-)} (\mathbf{r} ) = \int\limits_0^{\infty} d \omega \hat{\mathbf{B}}^{(\dagger)} (\mathbf{r},\omega)$
%%%
is the positive-frequency part, and the single component is given by 
\begin{eqnarray}\label{eq:B}
\hat{B}_{k} (\mathbf{r},\omega)&=& \sqrt{\frac{\hbar}{\pi \epsilon_{0}}} \frac{\omega}{c^2}
\int d^{3} \mathbf{r}' \sqrt{\epsilon_{I}(\mathbf{r}',\omega)} 
\nonumber \\
&&\epsilon_{kjs} \partial_{j} G_{sn} (\mathbf{r},\mathbf{r}',\omega) \hat{f}_{n}(\mathbf{ r}',\omega).
\end{eqnarray}
In the Heisenberg picture, the equation of motion for creation operator $\hat{a}_{j}^{+}$ is given by
\begin{eqnarray}\label{eq:adag}
\dot{\hat{a}}^{\dagger}_{j} (t)&=&i \omega_{v,j}\hat{a}^{\dagger}_{j}-  \frac{i}{\hbar} \frac{x_j}{2} 
\partial_j  \mu_k \hat{B}^{-}_{k}(\mathbf{r_{A}})\, , 
\end{eqnarray}
and the bosonic field operator becomes
\begin{eqnarray}
\lefteqn{
\dot{\hat{f}}_{n} (\mathbf{r},\omega, t)=-i \omega \hat{f}_{n}}
\\ &&
+i  \hat{a}_{j} \frac{ x_j}{2} \frac{\mu_k }{\sqrt{\hbar \pi \epsilon_{0}} } \frac{\omega}{c^2}
\sqrt{\epsilon_{I}(\mathbf{r},\omega)} \, 
  \partial_j  \epsilon_{kqs} \partial_{q} G^{*}_{sn} (\mathbf{r}_{A},\mathbf{r},\omega)\, , 
\nonumber
\end{eqnarray}
where the time integral in the Markov approximation becomes
%%%
\begin{eqnarray}\label{eq:fn}
\lefteqn{
\hat{f}_{n} (\mathbf{r},t)=
\hat{f}_{n,free}(\mathbf{r},t)+i \hat{a}_{j} (t)\zeta(\omega_{v,j}-\omega)}
\\&&
 \frac{x_j}{2} \frac{\mu_k  }{\sqrt{{\hbar \pi \epsilon_{0}}}}
\frac{\omega}{c^2}
\sqrt{\epsilon_{I}(\mathbf{r},\omega)}  \partial_j \epsilon_{kqs} \partial_{q} G^{*}_{sn} (\mathbf{r}_{A},\mathbf{r},\omega)\, , 
\nonumber
\end{eqnarray}
and $\zeta (x) = \pi \delta (x) + i \mathcal{P} x^{-1}$ ($\mathcal{P}$ denotes the principal part).  
This formal solution is going to be substituted into the magnetic field of Eq.~(\ref{eq:B})
and making use of the following integral relation for the Green function \cite{PERINA}
\begin{equation}
\label{magicformula}
\int d^3\mathbf{s} \frac{\omega^2}{c^2}
\epsilon_I(\mathbf{s},\omega) \bm{G}(\mathbf{r},\mathbf{s},\omega)
\bm{G}^*(\mathbf{r}',\mathbf{s},\omega) =
\mathrm{Im}\bm{G}(\mathbf{r},\mathbf{r}',\omega) \, ,
\end{equation}
we obtain
\begin{eqnarray}\label{eq:B2}
\lefteqn{
\hat{B}_{i} (\mathbf{r}_{A},\omega)= \hat{B}_{i,free} (\mathbf{r}_{A},\omega)
+ i  \hat{a}_{j} (t) \frac{x_j}{2} \frac{1}{ \pi \epsilon_{0} c^2}   }
\\ && \zeta(\omega_{v,j}-\omega)
\partial_j
\mathrm{Im} \left [\overrightarrow{\bm{\nabla}} \times  \bm{G}(\mathbf{r}_{A},\mathbf{r}_A,\omega) \times \overleftarrow{\bm{\nabla}} \right ]_{ik}  \mu_k\, , 
\nonumber
\end{eqnarray}
where the gradient $\overleftarrow{\bm{\nabla}}=\sum_{j=x,y,z} \partial_j $ is acting on the second argument of the Green function as denoted by the left pointing arrow. After integrating Eq.~(\ref{eq:B2}) over $\omega$ and substituting it into Eq.~(\ref{eq:adag}), we can write the equation of motion for the creation operator as
\begin{eqnarray}\label{eq:adagg2}
\lefteqn{
\dot{\hat{a}}^{\dagger}_{j} (t)
=\left [-\gamma_{j}/2 + i (\omega_{v,j} - \delta \omega_{j})\right ] \hat{a}^{\dagger}_{j}}
\nonumber \\ && 
-\frac{i}{\hbar} \frac{x_j}{2}
\partial_j \mu_{p} \hat{B}^{-}_{p,free}(\mathbf{r}_{A}). 
\end{eqnarray}
%%%%%%%%%%%%%%%%%%%%%%%%%%%%%%%%%
where $\gamma_{j}=\gamma_{j,n\rightarrow n\pm1}$ is the rate associated with the $n \rightarrow n\pm1$ transition along the $j$th direction. The rate  arises from the $\delta(x)$ of the $\zeta (x)$ function appearing in Eq.~(\ref{eq:fn}), and is given by 
\begin{eqnarray}\label{eq:HR}
\lefteqn{
\gamma_j =
\frac{ x_{j}^2 }{2\hbar \epsilon_0 c^2} }
\\
&& \overrightarrow{\bm{\nabla}} \left ( \bm{\mu} \cdot 
\mathrm{Im} \left [\overrightarrow{\bm{\nabla}} \times  \bm{G}(\mathbf{r}_{A},\mathbf{r}_A,\omega_{v,j}) \times \overleftarrow{\bm{\nabla}} \right ] \cdot  \bm{\mu} \right ) \overleftarrow{\bm{\nabla}}\, , 
\nonumber
\end{eqnarray}
while the term $\delta \omega_j$ of Eq.~(\ref{eq:adagg2}) arises from the principal-part of the $\zeta (x) $ function and denotes the frequency shift which we assume to be negligible.

If the dielectric body is in thermal equilibrium with its surroundings, then the magnetic field is in a thermal 
state with temperature $T$, and the magnetic field spectrum has to be multiplied by the mean thermal photon number $n_{\mathrm{th}}+1$. 
The spectrum of the magnetic field can be expressed as 
\begin{eqnarray}
\lefteqn{S_B (\mathbf{r}_A, \omega) = \frac{\hbar}{\pi \epsilon_0 c^2} (n_{\mathrm{th}}+1)} 
\\
&&\mathrm{Im} \left [\overrightarrow{\bm{\nabla}} \times  \bm{G}(\mathbf{r}_{A},\mathbf{r}_A,\omega) \times \overleftarrow{\bm{\nabla}} \right ]   \,, 
\nonumber
\end{eqnarray} 
and hence the spectrum of the fluctuating force causing the shift of the center of mass is 
\begin{eqnarray}
\lefteqn{
S_F (\mathbf{r}_A, \omega) = \frac{\hbar}{\pi \epsilon_0 c^2} (n_{\mathrm{th}}+1)}
\\
&& \overrightarrow{\bm{\nabla}} \left ( \bm{\mu} \cdot 
\mathrm{Im} \left [\overrightarrow{\bm{\nabla}} \times  \bm{G}(\mathbf{r}_{A},\mathbf{r}_A,\omega) \times \overleftarrow{\bm{\nabla}} \right ] \cdot  \bm{\mu} \right ) \overleftarrow{\bm{\nabla}}\, , 
\nonumber
\end{eqnarray} 
such that the transition rate for the $j$th trapping direction can be written as 
\begin{equation}
\gamma_{j,n \rightarrow n \pm 1}=\frac{\pi x_{j,\pm}^2}{2\hbar^2} S_F ( \mathbf{r}_A, \omega_{v,j}).\end{equation}

\end{document}